\documentclass[%
 reprint,
superscriptaddress,
 amsmath,amssymb,
 aps,
]{revtex4-1}

\usepackage{graphicx}
\usepackage{dcolumn}
\usepackage{bm}
\usepackage{color}

\begin{document}

\preprint{APS/123-QED}

\title{On-Demand Microwave Generator of Shaped Single Photons}

\author{P.~Forn-D\'iaz}
\email{pol.forndiaz@bsc.es}
\affiliation{%
 Institute for Quantum Computing and Department of Electrical and Computer Engineering, University of Waterloo, 200 University Avenue West, Waterloo, N2L 3G1, Canada
}%
\affiliation{Escola T\`ecnica Superior d'Enginyeria Industrial de Barcelona, Universitat Polit\`ecnica de Catalunya, Barcelona, 08034, Spain}
\affiliation{%
 Barcelona Supercomputing Center (BSC), Carrer de Jordi Girona 29, 08034, Barcelona. Spain
}%
\author{C.~W.~Warren}%
\affiliation{%
 Institute for Quantum Computing and Department of Electrical and Computer Engineering, University of Waterloo, 200 University Avenue West, Waterloo, N2L 3G1, Canada
}%
\author{C.~W.~S.~Chang}%
\affiliation{%
 Institute for Quantum Computing and Department of Electrical and Computer Engineering, University of Waterloo, 200 University Avenue West, Waterloo, N2L 3G1, Canada
}%
\author{A.~M.~Vadiraj}%
\affiliation{%
 Institute for Quantum Computing and Department of Electrical and Computer Engineering, University of Waterloo, 200 University Avenue West, Waterloo, N2L 3G1, Canada
}
\author{C.~M.~Wilson}%
\email{chris.wilson@uwaterloo.ca}
\affiliation{%
 Institute for Quantum Computing and Department of Electrical and Computer Engineering, University of Waterloo, 200 University Avenue West, Waterloo, N2L 3G1, Canada
}%

\date{\today}

\begin{abstract}
We demonstrate the full functionality of a circuit that generates single microwave photons on demand,
with a wave packet that can be modulated with a near-arbitrary shape. We achieve such a high tunability by
coupling a superconducting qubit near the end of a semi-infinite transmission line. A dc superconducting
quantum interference device shunts the line to ground and is employed to modify the spatial dependence of
the electromagnetic mode structure in the transmission line. This control allows us to couple and decouple
the qubit from the line, shaping its emission rate on fast time scales. Our decoupling scheme is applicable to
all types of superconducting qubits and other solid-state systems and can be generalized to multiple qubits
as well as to resonators.

\end{abstract}

\pacs{Valid PACS appear here}
                             
\maketitle

\section{\label{sec:intro}Introduction}
Quantum networks are a promising paradigm both for quantum communication and quantum computation. A quantum network consists of a set of quantum processing nodes, between which quantum information is communicated using quantum channels \cite{kimble2008}. Individual photons are promising information carriers between quantum nodes given the robustness of their quantum states and the possibility of traveling long distances, both in the optical \cite{hensen2015} and the microwave frequency domain \cite{vermersch-xiang2017}. For quantum networks and other purposes, single-photon generators are therefore an enabling resource. A list of desirable characteristics for a single-photon source include high-efficiency, on-demand (deterministic) operation, and low timing jitter in the emission. In order to achieve maximum state transfer efficiency between quantum nodes, it is also desirable to controllably shape the photon wave packet, such that it matches the absorption profile of the receiving node \cite{cirac1997}. Single-photon sources have other applications beyond quantum communication, such as preparing nonclassical states of harmonic oscillators which can be used in certain quantum error-correcting protocols \cite{ofek2016}.

Superconducting qubits are ideal candidates to behave as tunable single-photon sources in the microwave domain. There have already been implementations of single-photon sources made from qubits coupled to resonators \cite{houck2007, bozyigit2011, eichler2011, kindel2016}, including circuits with tunable waveforms \cite{pechal2014, yin2013, wenner2014}. While shown to be useful for specific applications such as Hong-Ou-Mandel interferometry \cite{hong1987, lang2013}, the usage of a cavity in these implementations intrinsically limits the source to a single frequency. Working instead with qubits directly coupled to propagating modes of radiation enables the engineering of tunable, highly efficient single-photon sources \cite{roy2017}. In recent years, experiments have demonstrated basic operations of a single qubit interacting with propagating modes such as resonance fluorescence \cite{astafiev2010}, a single-photon router \cite{hoi2011}, quantum-state filtering \cite{hoi2012}, and photon-mediated interactions \cite{hoi2013, vanloo2013}. Recent experiments using transmission lines have demonstrated a frequency-tunable single-photon source with a flux qubit capacitively coupled to two semi-infinite lines \cite{peng2016} and the correlated emission of photons at two different frequencies using a transmon qubit \cite{gasparinetti2017}. Both experiments used fixed qubit-line couplings. Despite all of the progress, no experiment to date has demonstrated a qubit-line interaction that can be controllably switched off, an important advantage in efficiently transferring quantum states between remote nodes of a quantum network \cite{kimble2008}. Recent theoretical proposals have suggested the use of a dc superconducting quantum interference device as an adjustable impedance to modify the spatial profile of vacuum fluctuations along a semi-infinite transmission line \cite{koshino2012, sankar2016}, thereby controlling the photon emission rate of the qubit. The basic principle of this tunable-coupling scheme was demonstrated in an earlier experiment \cite{hoi2015}, but only with static tuning. Besides the relevant applicability as a tunable single-photon source, a system with a tunable interaction to a continuum of modes would allow fundamental studies of light-matter interaction including generalizations of the Lamb shift \cite{lamb1947}. 

Here, we implement a superconducting circuit that generates shaped single microwave photons on demand, with low timing jitter and high efficiency. Our device consists of a dc SQUID operating as a tunable shunt in a semi-infinite transmission line, together with a capacitively coupled transmon qubit \cite{koch2007} positioned a specific distance from the end of the line. The relative simplicity of the device, compared to its high performance, is enabled by the versatility and novelty of our tunable-coupling scheme.  

\section{Principle of operation}\label{sec2}
Our circuit consists of a semi-infinite transmission line, the \emph{source} line, with a dc SQUID acting as an inductive shunt embedded at its end [see Fig.~\ref{fig1}(a)]. The impedance presented by the SQUID can be tuned with the application of an external magnetic flux $\Phi_{\rm SQ}$. In this configuration, the SQUID effectively behaves as an electrical mirror with an adjustable position, which is complementary to a recent experiment where a qubit was instead effectively moved by tuning its frequency \cite{hoi2015}. We galvanically connect a high-bandwidth transmission line, the \emph{current} line, to the SQUID loop to enable fast tuning of $\Phi_{\rm SQ}$. Using galvanic coupling maximizes the tuning for a given amount of current, minimizing the amount of dissipation produced by the applied current pulses. A superconducting transmon qubit \cite{koch2007} is capacitively coupled at a specific distance from the end of the source line. A separate on-chip microwave line, the \emph{excitation} line, is weakly coupled to the qubit, enabling direct driving of the qubit even when it is decoupled from the source line.

Ideally, the SQUID would present no impedance when biased at zero flux, $Z_{\rm SQ}(0) = 0$, and would become an open circuit at $\Phi_{\rm SQ} = \Phi_0/2$ that is, $Z_{\rm}(\Phi_0/2)\to\infty$. In this case, the flux-dependent reflection coefficient of the SQUID, $\Gamma_{\rm SQ} = \exp[{i\phi(\Phi_{\rm SQ})}]$, would change its phase from $\phi(0) = \pi$ to $\phi(\Phi_0/2) = 0$ \cite{Pozar}. In practice, the finite impedance presented by the SQUID results in a phase shift given by
\begin{equation}\label{eq:phi}
\tan[\phi(\Phi_{\rm SQ})] = \frac{2\frac{Z_0}{\omega L_{\rm SQ}}\bigg[1-\Big(\frac{\omega}{\omega_{\rm SQ}}\Big)^2\bigg]}{1- \left(\frac{Z_0}{\omega L_{\rm SQ}}\right)^2\bigg[1-\Big(\frac{\omega}{\omega_{\rm SQ}}\Big)^2\bigg]^{2}}.
\end{equation}
Here, $Z_0$ is the impedance of the transmission line, $L_{\rm SQ} \equiv\Phi_0[2\pi I_C|\cos(\pi\Phi_{\rm SQ}/\Phi_0)|]^{-1}$ is the SQUID Josephson inductance, and $\omega_{\rm SQ} = (L_{\rm SQ}C_{\rm SQ})^{-1/2}$ is the SQUID plasma frequency, with $C_{\rm SQ}$ being the total SQUID capacitance. Equation~(\ref{eq:phi}) contains the ideal cases of a perfect short, $L_{\rm SQ}=0$, and a perfect open circuit, $L_{\rm SQ}=\infty$. We can reexpress the total phase $\phi$ by separating the $\pi$ phase of an ideal short circuit from the additional phase due to the finite SQUID impedance, $\phi = \pi - \phi_{\rm SQ}$. The phase is negative since we assume the line extends towards $x=-\infty$, as represented in Fig.~\ref{fig1}(a). Following from Ref.~\cite{Johansson2009}, the additional phase $\phi_{\rm SQ}$ can be effectively recast as an additional electrical length of the line $l_{\rm eff}$, with 
\begin{equation}\label{eq:Gamma}
\phi_{\rm SQ} = 2kl_{\rm eff},
\end{equation}
where $k = \omega/v$ is the wave number of a mode with frequency $\omega$ and $v$ is the speed of light in the transmission line. Equation~(\ref{eq:Gamma}) shows how a SQUID can be used to change the effective length of a transmission line \cite{Wilson2011}. In the limit of small SQUID inductance, $\phi_{\rm SQ} \simeq 2\omega L_{\rm SQ}/Z_0$ and $l_{\rm eff} \simeq L_{\rm SQ}/L_0$, as in Ref.~\cite{Johansson2009}, with $L_0$ being the inductance per unit length of the line. 

At a distance $x$ from the end of the line, the round-trip phase acquired by a wave with frequency $\omega=2\pi v/\lambda$ propagating to the SQUID and back again is $\theta_T(\Phi_{\rm SQ},\omega,x) \equiv\pi -\phi_{\rm SQ} - \phi_L(x)$, with $\phi_L \equiv 4\pi x/\lambda = 2\omega x/v$. Owing to destructive interference, the voltage amplitude of a standing-wave mode will vanish whenever $\theta_T(\Phi_{\rm SQ},\omega,x)=\pi$.  Crucially, this statement applies equally well to vacuum modes and classical modes. A qubit with frequency $\omega_{01}$ positioned at $x$ will therefore be effectively decoupled from the line when $\theta_T(\Phi_{\rm SQ},\omega_{01},x)=\pi$. Defining the qubit wavelength $\lambda_{01} = 2\pi v/\omega_{01}$, the decoupling point will be near $\Phi_{\rm SQ} = 0$ for $x=\lambda_{01}/2$ or near $\Phi = \Phi_0/2$ for $x = \lambda_{01}/4$. In the ideal case where Eq.~(\ref{eq:phi}) changes from $\pi$ to 0 as the SQUID flux changes from 0 to $\Phi_0/2$, the electrical length of the line effectively increases by $\lambda/4$. However, the finite impedance of a real SQUID will reduce the total phase shift, reducing the dynamic range of the qubit-line coupling.

The interference just described also shapes the power spectral density of the vacuum modes \cite{hoi2015}. It becomes
\begin{equation}\label{eq:S}
S(\omega) = 2\hbar\omega\cos^2\bigg(\frac{\pi}{2} - 2\pi\frac{x+l_{\rm eff}}{\lambda}\bigg).
\end{equation}
By changing $l_{\rm eff}$ using the SQUID flux, we can therefore change the vacuum spectral density seen by the qubit. 

To understand how modifying the vacuum mode structure changes the effective coupling, we recall that any quantum emitter coupled to a continuum of modes will experience energy decay due to vacuum fluctuations of the continuum. Following from Fermi's ``golden rule", the emission rate, $\Gamma_1$, is proportional to the spectral density of modes at the emitter frequency, $\Gamma_1\sim S(\omega_{01})$. It was shown \cite{hoi2015} that, for a transmon qubit in front of a mirror, the precise relation is 
\begin{equation}\label{eq:G1}
\Gamma_1 = Z_0\left(\frac{e}{\hbar}\right)^2\left(\frac{C_s}{C_{\Sigma}}\right)^2\sqrt{\frac{E_J}{2E_C}}S(\omega_{01}). 
\end{equation}
Here, $C_s$ is the qubit-line capacitance, and $C_{\Sigma} = C_s + C_J + C_e$ is the total capacitance including the qubit self-capacitance $C_J$ and the capacitance to control lines $C_e$. $E_J$ is the qubit Josephson energy and $E_C$ is the qubit charging energy. With $S(\omega_{01})$ in the source line given by Eq.~(\ref{eq:S}), we see that the emission rate into the line can be tuned by adjusting $l_{\rm eff}$. In this work, we use this effect to continuously tune $\Gamma_1$ on short time scales by adjusting $\Phi_{\rm SQ}$.

In order to build an on-demand photon generator, we first apply a magnetic flux through the SQUID that sets $S(\omega_{01}) = 0$ exactly at the qubit position [the blue curve in Fig.~\ref{fig1}(a)], suppressing qubit emission into the line. The qubit can then be prepared in an arbitrary state by exciting it using the excitation line, which is always coupled. Using the fast current line, a current pulse of arbitrary shape modifies $\Phi_{\rm SQ}(t)$ in real time, which translates into a shift of the position of vacuum field nodes. The qubit is then coupled to the vacuum fluctuations [the red curve in Fig.~\ref{fig1}(a)] and spontaneously emits a photon which contains the information of the qubit state in its in-phase ($I$) and quadrature ($Q$) components \cite{bozyigit2011}. As the qubit-line coupling can be controlled in real time by the SQUID current pulse, the emitted photons can be shaped arbitrarily.

In our experiment, we use a single-junction, fixed-frequency transmon to keep $\omega_{01}$ fixed. We could extend our circuit's capabilities to allow tuning of the frequency of the emitted photons by using a standard tunable-frequency transmon without modifying any of the protocols presented in this work. 

\section{Device description}
The device is fabricated in three lithography steps. In the first step, Pd $e$-beam markers are patterned using UV lithography, followed by an evaporation of 5~nm of Ti  and 70~nm of Pd. We use Pd instead of the more typical Au to avoid interdiffusion with Al. The first Al layer containing the transmission line and control lines is patterned with UV lithography. The thickness of the evaporated aluminum is 100~nm. The final lithography step contains the transmon qubit and the SQUID, which is patterned using $e$-beam lithography. The evaporation consists of two layers of 40 and 60 nm, with an oxidation step in between with a pressure of 25~Torr for 15 min. \emph {In-situ} Ar milling is used prior to the junction evaporation to guarantee good galvanic contact between the layers.

\begin{figure}[!hbt]
\centering
\includegraphics{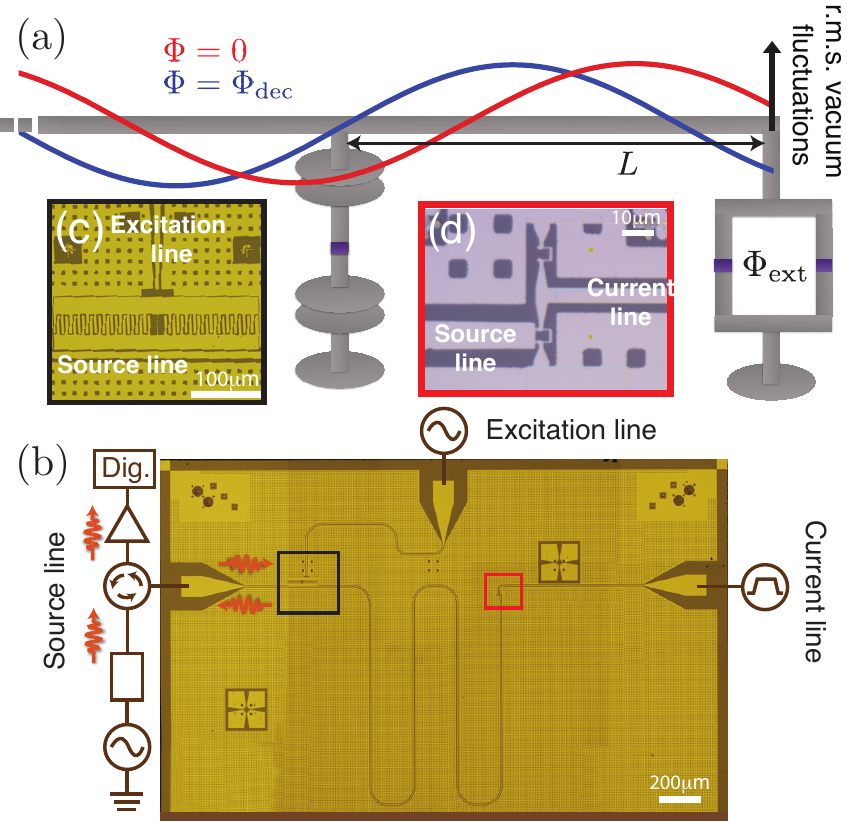}
\caption{\label{fig1} (a) Sketch of the device operating principle. A dc SQUID shunts the line to ground and a flux threading it controls the profile of the rms~amplitude of the vacuum fluctuations. The field offset at $\Phi = 0$ is due to the finite SQUID impedance. We plot a single mode for clarity, but the effect is broadband. A qubit of frequency $\omega_{01}=2\pi v/\lambda_{01}$ is positioned near an odd number of quarter wavelengths, $L=(2n+1)\lambda_{01}/4$. At $\Phi=0$, depicted in red, the qubit is maximally coupled to the line, even though the field offset lowers the maximum coupling as a function of the qubit wavelength. At a particular decoupling flux $\Phi=\Phi_{\rm dec}$, depicted in blue, the qubit is decoupled from the line. (b) Optical image of a device similar to the one used in the experiment and the circuit schematic. The output from the source line is amplified at different stages and digitized at room temperature. (c) Optical image of the single-junction transmon qubit capacitively coupled to the transmission line. The excitation line has a weak, fixed coupling to the qubit, allowing its state to be controlled irrespective of the coupling strength to the source line. (d) Optical image of the SQUID shunting the source line to ground. The current line connecting from the left is used to tune the SQUID flux $\Phi_{\rm SQ}$ in real time.}
\end{figure}

A micrograph of a device similar to the one measured is shown in Fig.~\ref{fig1}(b). The transmon qubit capacitively couples to the source line over a length of 300~$\mu\rm{m}$ with an estimated capacitance of $C_s \approx28~\rm{fF}$
 [see Fig.~\ref{fig1}(c)]. The distance from the center of the qubit capacitor to the end of the line is $L=9.59$~mm. The excitation line is weakly coupled to one of the qubit capacitor plates with $C_e\approx0.9~\rm{fF}$. The current line galvanically coupled to the SQUID loop is symmetrically split [see Fig.~\ref{fig1}(d)]. The mutual inductance of the SQUID-bias line is estimated to be 4~pH. 

Qubit spectroscopy can be performed directly by measuring the reflection of a probing field [see Fig.~\ref{fig1}(b)]. Alternatively, the qubit spectrum can be measured in transmission by using the excitation line as the input port. Using both methods, we measure a qubit frequency of $\omega_{01}/2\pi = 3.690~$GHz with anharmonicity $\alpha/2\pi = -141.7~$MHz, yielding a total qubit capacitance of $C_{\Sigma}=136~\rm{fF}$, $E_J/E_C = 91$, $E_J/h = 12.9~$GHz, and a qubit critical current of $I_C=26~$nA. 

We spectroscopically characterize the system by measuring the reflection of a coherent signal off the circuit with the reflection coefficient $r\equiv \langle V_r\rangle/\langle V_{\rm in}\rangle$, where $\langle V_r\rangle$ and $\langle V_{\rm in}\rangle$ are the phase-sensitive averages of the reflected and incident voltages, respectively. In Ref.~\cite{hoi2015}, it was shown that, for a weak probe,
\begin{equation}\label{eq2}
r = -1 + \frac{\Gamma_1}{\Gamma_2 + i\delta\omega},
\end{equation}
with $\Gamma_2 = \Gamma_1/2 + \Gamma_{\phi}$ being the total decoherence rate and $\Gamma_{\phi}$ the pure dephasing rate. $\delta\omega = \omega_{01} - \omega_d$ is the drive detuning from the qubit frequency $\omega_{01}$. $\Gamma_1$ is the emission rate of the qubit into the transmission line and therefore quantifies the qubit-line coupling strength. 

\begin{figure}[!hbt]
\centering
\includegraphics{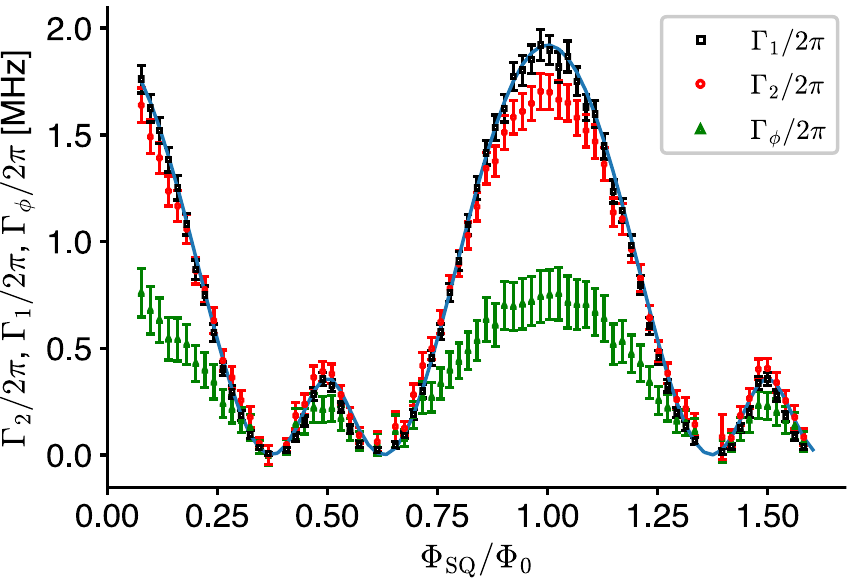}
\caption{\label{fig2}Extracted qubit decay rates. The qubit emission rate into the source line, $\Gamma_1$ (the black squares), and the total decoherence rate, $\Gamma_2$ (the red circles), are obtained from Eq.~(\ref{eq2}). The pure dephasing rate $\Gamma_{\phi}$ is calculated as $\Gamma_{\phi} = \Gamma_2 - \Gamma_1/2$. All three rates tune with flux, showing a clear modulation of the qubit--source line coupling strength. The blue trace is a fit using Eqs.~(\ref{eq:phi}), (\ref{eq:G1}), and (\ref{eq:mismatch}), including SQUID junction asymmetry.}
\end{figure}

Figure~\ref{fig2} shows qubit decay rates extracted from spectroscopy measured in reflection, as a function of $\Phi_{\rm SQ}$. When $\Phi_{\rm SQ}$ is an integer multiple of $\Phi_0$, the qubit exhibits a maximum decay rate $\Gamma_1/2\pi = 1.9~$MHz. Within a flux period, there are two points where the decay rate is observed to approximately vanish, $\Phi_{\rm SQ}^{(1)}/\Phi_0 = 0.39$ and $\Phi_{\rm SQ}^{(2)}/\Phi_0 = 0.61$. At these flux values, the qubit signal is lost, indicating a clear decoupling from the source line. At these points, the driving field does not excite the qubit and the signal is fully reflected. The modulation of $\Gamma_1$ is a signature of the tunability of the qubit-line interaction.
The observed maximum and minimum of $\Gamma_1$ allow us to place a bound on the \emph{on} or \emph{off} ratio of the decoupling mechanism at approximately 35. The qubit frequency is also modulated with the external flux by a small amount (not shown), even though this is a single-junction transmon qubit. This frequency shift will be the study of future work \cite{forn-diazX}. 

According to Eq.~(\ref{eq:G1}), the maximum emission rate using the estimated $C_s$ should be 26~MHz. As explained in Sec.~\ref{sec2}, the maximum tuning range of the device is limited by the finite SQUID impedance. For this reason, we are not able to tune the device to this ideal maximum coupling rate. 

The fact that the maximum observed emission rate occurs near an integer number of flux quanta (Fig.~\ref{fig2}) indicates that the length of the line is close to $\lambda_{01}/4$. The observed recovery of the emission rate at $\Phi_{\rm SQ}/\Phi_0 = 0.5$ permits us to calculate the total length of the circuit. The decoupling point lies 0.11$\Phi_0$ away from $\Phi_0/2$, which corresponds to 22\% of $\Phi_0/2$. Therefore, the length of the line is $L = \lambda_{01}/4 + 0.22\lambda_{01}/4 = 9.49~$mm, which is within 100~$\mu\rm{m}$ of the designed value. The fact that the emission rate vanishes in Fig.~\ref{fig2} is consistent with a positive length offset. With a negative length offset, the emission rate would never vanish. The mismatch in length from $\lambda_{01}/4$ reduces the dynamic range of the maximum emission rate $\Gamma_1$ even further. In a future device, this mismatch could be circumvented with a frequency-tunable qubit. 

The form of the power spectral density can be modified to accommodate for the length mismatch with a phase offset $\phi_{\rm off}$. Setting $\phi_L/2 = (\lambda/4)(4\pi/\lambda)/2 + \phi_{\rm off}= \pi/2 + \phi_{\rm off}$ in Eq.~(\ref{eq:S}), the power spectral density becomes
\begin{equation}\label{eq:mismatch}
S(\omega) = 2\hbar\omega\cos^2(\phi_{\rm SQ}/2 -\phi_{\rm off}).
\end{equation}
Using Eqs.~(\ref{eq:phi}), (\ref{eq:G1}), and (\ref{eq:mismatch}) we can very accurately fit the emission rate (the blue curve in Fig.~{\ref{fig2}}) to extract the device parameters influencing the dynamic range of $\Gamma_1$. The fit requires introducing asymmetry between the two SQUID junctions, which is typical in real SQUIDs. The extracted critical currents of the SQUID junctions are $I_{C1} = 30\pm1~\rm{nA}$ and $I_{C2} = 46\pm1~\rm{nA}$. We also extract $C_{\rm SQ} = 33\pm1~\rm{fF}$. The qubit-line coupling capacitance is $C_c = 31.0\pm1.2~\rm{fF}$, which is close to the calculated value. The reduced dynamic range of the emission rate can then be fully explained by the finite SQUID impedance, the SQUID junction asymmetry, and line length mismatch. Using a larger critical current junction and a tunable-frequency qubit allows us to increase the dynamic range and, therefore, the \emph{on} or \emph{off} ratio of the photon generator.

The pure dephasing rate $\Gamma_{\phi}$ clearly modulates along with the qubit-line coupling until saturating at a value of $\Gamma_{\phi}/2\pi\approx0.7~\rm{MHz}$ in the range of flux $\Phi_{\rm SQ}/\Phi_0 = 0.9-1.1$. We do not currently have a clear explanation for this dependence. 

\section{Driven-qubit dynamics}
We can also see the tunability of the qubit--source line coupling in the real-time dynamics of the qubit when a coherent microwave control pulse is applied. Our detected signals correspond to the quadratures of the field emitted by the qubit $\langle I(t)\rangle, \langle Q(t)\rangle$, where the expectation value is evaluated for the qubit state at the time of emission. Each individual time trace is typically averaged $10^6$ times. As was already demonstrated in previous work using a transmon qubit coupled to a resonator \cite{houck2007, bozyigit2011}, the qubit coherence is mapped to the state of the emitted photon. Following the standard result from input-output theory \cite{houck2007}, given the existing direct relationship between the qubit emission operator $\langle\sigma^-\rangle$ and the quadratures of the emitted field $\langle (a\pm a^{\dag})\rangle$ \cite{abdum2011}, a maximum of the recorded signal corresponds to the qubit in a superposition state on the equatorial plane of the Bloch sphere, such as $|\pm\rangle = (|0\rangle\pm|1\rangle)/\sqrt{2}$.
\begin{figure}[!hb]
\centering
\includegraphics{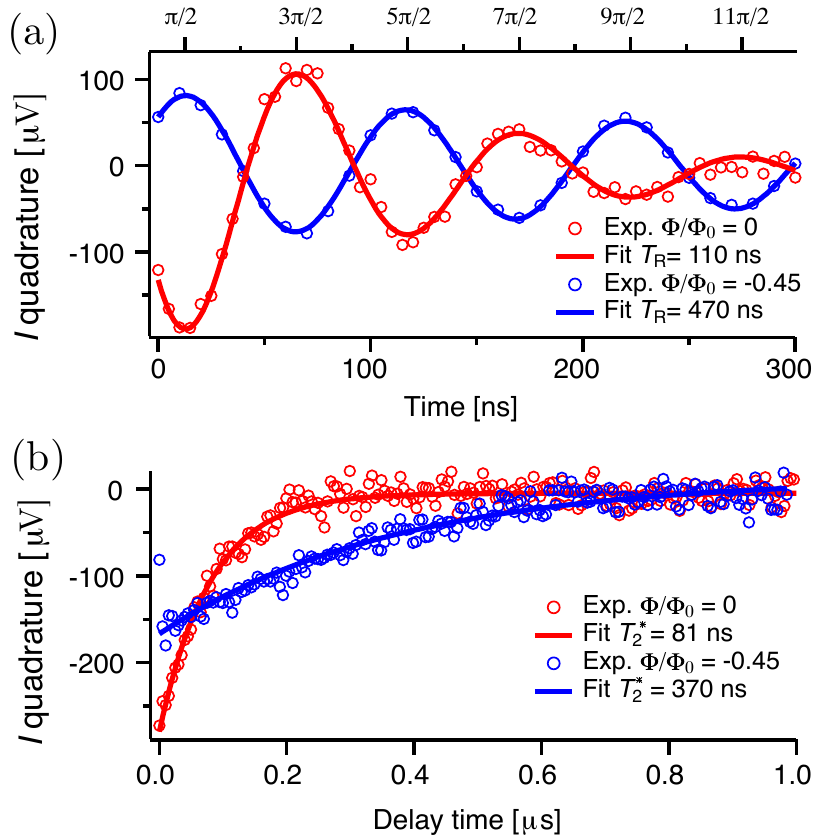}
\caption{\label{fig3}Qubit dynamics at two operating flux bias points with higher (red) and lower (blue) qubit--source line coupling rate. (a) Rabi oscillations. The Rabi decay times, $T_R$, differ by a factor of 4.3. The upper axis indicates the angle of the qubit-state rotation on the Bloch sphere. (b) Free induction decay of the qubit state after a $\pi/2$ pulse. The decay rate is enhanced 4.6 times between the two bias points. The enhanced coupling of the qubit into the line is further evidenced by a larger qubit signal for the shortest-lived traces, for both the free decay and the Rabi oscillations.}
\end{figure}

Figure~\ref{fig3}(a) shows Rabi oscillations of the qubit biased at two different bias points with significantly different coupling to the source line, while Fig.~\ref{fig3}(b) shows the free decay of the qubit after a $\pi/2$ pulse. The Rabi oscillations are obtained by integrating free-decay traces such as those in Fig.~\ref{fig3}(b) at different Rabi pulse lengths. The integrated value is then normalized to the number of points. In all traces in Fig.~3, the signal is maximized by digitally rotating it into the $I$ quadrature. The red data points in Figs.~\ref{fig3}(a) and \ref{fig3}(b) correspond to $\Phi_{\rm{SQ}}/\Phi_0=0$. The observed Rabi oscillations at $\Phi_{\rm SQ}=0$ in Fig.~\ref{fig3}(a) decay with a characteristic time of $T_R=110~$ns. The emitted qubit signal is maximum after a $\pi/2$ pulse since we are recording the qubit quadratures and therefore the qubit coherence $\langle\sigma^-\rangle$ as explained above. The qubit free decay in Fig.~\ref{fig3}(b) at $\Phi_{\rm SQ}=0$ is $T_2^{\ast}=81~$ns when prepared in a superposition state. This decay time corresponds to a decay rate of $\Gamma_2/2\pi = 1.96~\rm{MHz}$, which is consistent with the spectroscopic value of $\Gamma_2$ in Fig.~2. The discrepancy between the two measurements is likely due to phase drift between the microwave generators used to record the time-domain data that appear as additional dephasing, or fluctuations in $T_1$. The blue data points in Figs.~\ref{fig3}(a) and \ref{fig3}(b) correspond to $\Phi_{\rm{SQ}}/\Phi_0 = -0.45$. The Rabi oscillations observed in Fig.~\ref{fig3}(a) are clearly longer lived compared to those at $\Phi_{\rm SQ}/\Phi_0=0$, indicating a lower qubit--source line coupling. The decay of the Rabi oscillations at this bias point is $T_R=470~$ns, about 4 times longer than the decay at $\Phi_{\rm SQ}/\Phi_0 = 0$. The free decay of the qubit signal in Fig.~\ref{fig3}(b) is measured to be $T_2^{\ast}=370$~ns. In both Figs.~\ref{fig3}(a) and Fig.~\ref{fig3}(b), the amplitude of the qubit signal at $\Phi_{\rm SQ}/\Phi_0 = -0.45$ is lower than the signal at $\Phi_{\rm SQ}/\Phi_0=0$. This is another clear indication of the modulation of the qubit--source line coupling strength. 

\begin{figure}[!hbt]
\centering
\includegraphics{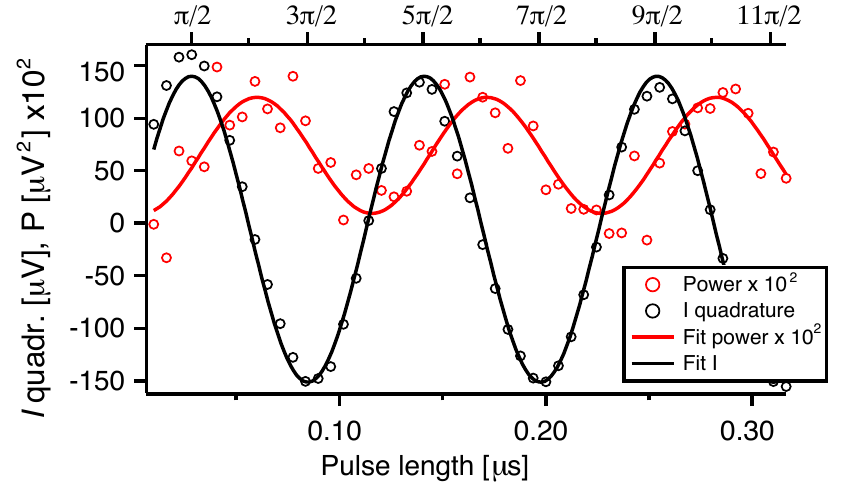}
\caption{\label{fig4}Power and quadrature oscillations of emitted radiation from the driven qubit. Black circles represent the measured $I$ quadrature amplitude, while red circles correspond to the second moment of the emission, the emitted power $\langle P\rangle=\langle(I^2+Q^2)\rangle$. The solid curves are sinusoidal fits to the data. The two curves are offset by $(0.58 \pm 0.08)\pi$, which rules out a coherent state and provides strong evidence for single-photon emission. The upper axis indicates the angle of the qubit-state rotation on the Bloch sphere.}
\end{figure}

To demonstrate the nonclassical nature of the emitted radiation, we record the second moment of the emitted field which corresponds to the emitted power $\langle P \rangle \equiv \langle (I^2+Q^2)\rangle$. Figure~\ref{fig4} shows the evolution of $\langle I\rangle$ and $\langle P\rangle$ during Rabi oscillations. Two traces are shown with the first (black) and second (red) moments of the driven qubit biased at $\Phi_{\rm{SQ}}/\Phi_0 = -0.45$, where the decay time is long. The signal is averaged over $2.5\times10^7$ repetitions, as $\langle P\rangle$ is rather noisy. Each data point of the power oscillation is the result of subtracting a background offset which fluctuates from trace to trace. The clear oscillatory pattern recovered after all of those averages is a manifestation of the coherence of the radiation emitted by the qubit. The phase of the power oscillation is clearly offset by approximately $\pi/2$ with respect to the quadrature amplitude oscillation, as is expected for a single-photon emitter such as our transmon qubit. For a coherent state, one would expect $\langle I_c^2\rangle = \langle I_c\rangle^2$; therefore, power and amplitude should oscillate in phase. By contrast, for the qubit state $\cos(\theta/2)|0\rangle + \sin(\theta/2)|1\rangle$, we get \cite{abdum2011} $\langle I_q\rangle\sim\langle\sigma^-\rangle \sim \sin\theta$ and $\langle P_q\rangle\sim1+\langle\sigma_z\rangle\sim 1-\cos\theta$, consistent with the observations in Fig.~\ref{fig4}. This is strong evidence that the source of photons is nonclassical \cite{houck2007}. 

\section{Triggered photon generation}
As mentioned in the Introduction, triggered photon emission is important for transferring quantum states in a quantum network with high efficiency. 

In order to demonstrate triggered photon emission with our circuit, we add a current pulse through the current line [see Fig.~\ref{fig1}(b)]. The current pulse tunes $\phi_{\rm SQ}(\Phi_{\rm SQ})$ in Eq.~(\ref{eq:phi}), allowing fast tuning of the qubit--source line coupling strength. The external field is first set to zero, $\Phi_{\rm{SQ}}= 0$, where the qubit is strongly coupled to the line, as shown in Fig.~\ref{fig2}. Figure~5 shows a schematic of the pulse scheme. A square current pulse is applied to bring the qubit to the decoupling bias point. Then, a pulse in the excitation line energizes the source. (For these experiments, we use a $\pi/2$ pulse to maximize the emitted quadrature signal, although a $\pi$ pulse would be used to produce a pure single photon.) The excitation can now be stored for a maximum of the intrinsic relaxation time of the qubit $T_{1,i}$. At a desired time, the current pulse returns the qubit to a strongly coupled point, where the excitation is emitted with a ``jitter" corresponding to the emission time of the qubit $T_1$ at that bias point. 

\begin{figure}[!hbt]
\centering
\includegraphics{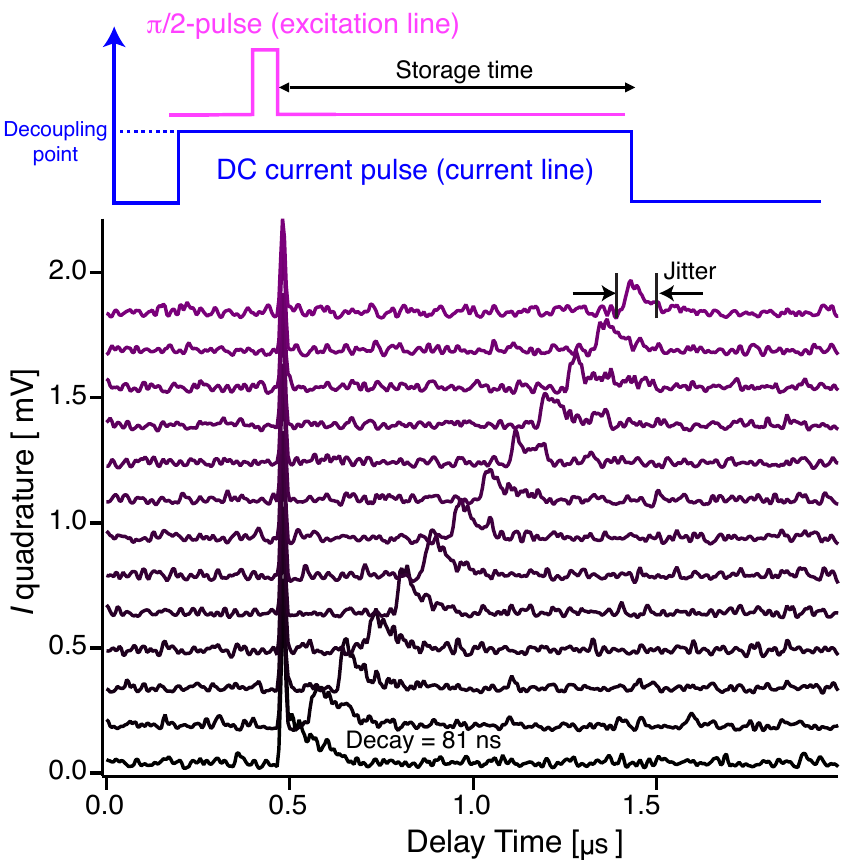}
\caption{\label{fig5} On-demand single-photon generation. The pulse sequence begins by bringing the qubit to the decoupling point by pulsing current through the SQUID and then exciting the qubit with a $\pi/2$ pulse. The qubit emission is triggered by ending the current pulse through the SQUID, bringing the qubit back to a strongly coupled point. The peak registered near 0.5~$\mu$s is the excitation pulse partially leaking into the measurement chain circuit.}
\end{figure}

Figure~\ref{fig5} shows the observed emission pattern following the pulse scheme just described. Each trace corresponds to a different storage time. The traces are vertically displaced for clarity. At a delay time of 0.5$~\mu\rm{s}$, a peak is always detected which corresponds to the excitation pulse leaking into the source line. The first trace shows a decaying tail of 81~ns immediately after the driven pulse. This signal corresponds to the emission of this particular device when it is excited at the bias point with maximum qubit coupling to the line. The emission pattern is similar to previous observations of single-photon sources with fixed coupling \cite{houck2007, peng2016, gasparinetti2017}. In the rest of the traces, the qubit excitation is stored for a controlled amount of time. The emission of the qubit can still be detected with delay times of up to $2~\mu\rm{s}$ for this particular device, demonstrating a triggered emission of photons on demand. 

The ratio of the storage time and the jitter time is a figure of merit for an ``on-demand" source. First, it is useful to be able to separate in time the excitation pulse and the single-photon emission, e.g., because the excitation pulse can blind detectors. Second, in applications where photon timing is important, the emission time of the photon $T_1$ represents a jitter (uncertainty) in the effective emission time. Minimizing the jitter at fixed coupling, however, limits the preparation fidelity, as the qubit partially decays during the excitation pulse. The minimization of jitter time in combination with a long storage time is therefore an important advantage of our photon generator over other implementations. 

In future work, the scheme is to be combined with a frequency-tunable qubit. This combination will allow the color of the emitted photons to be changed without compromising the emission efficiency, as is the case in other proposed single-photon source schemes based on qubits in cavities.

\section{Shaped-photon generation}\label{sec6}
The qubit emission can be further controlled in real time by sending shaped current pulses through the current line. In this way, the qubit emission rate $\Gamma_1[\Phi_{\rm{SQ}}(t)]$ can be adjusted nearly arbitrarily. Recent theoretical work \cite{gough2012, sankar2016} studied how the controlled emission then produces a known photon wave packet with amplitude $\xi(t)$. 
\begin{figure}[!hbt]
\centering
\includegraphics{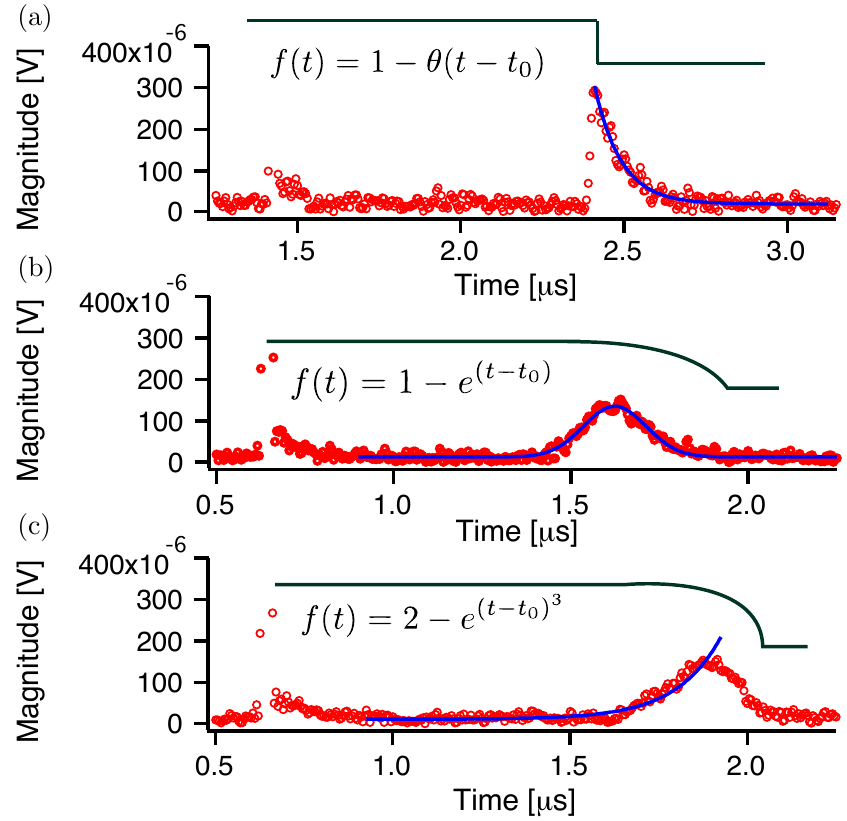}
\caption{\label{fig6}Shaped-photon generation. (a)--(c) Emission of the source given differing current modulations $f(t)$, sketched with the black traces above the data. The solid blue lines represent the targeted waveforms. (a) Emitted qubit voltage with a sharp falling edge. The qubit emission is clearly exponential. (b) Same as (a) but using a falling exponential edge. The qubit emission looks to be nearly symmetric. (c) The falling edge of the SQUID current pulse is an inverted cubic exponential. The qubit emission in this case approaches the time reverse of the emission in (a). The signal recorded $1~\mu\rm{s}$ prior to the qubit emission corresponds to the excitation pulse.} 
\end{figure}
Figure~\ref{fig6} displays several examples of measured shaped-photon emission. Figure \ref{fig6}(a) shows the exponentially decaying signal from a sharp-edged dc SQUID current pulse, which is the same profile observed in Fig.~\ref{fig5}. Figure \ref{fig6}(b) displays a nearly symmetric signal achieved by modulating the trailing edge of the dc pulse by an inverted exponential. Figure \ref{fig6}(c) shows the emission from an inverted cubic exponential which resembles a time-reversed version from the original exponential pulse in figure \ref{fig6}(a). Our electronics are limited in this experiment, so we simply adjust the parameters of the function generator by hand to produce the best-looking output waveform. However, with a more sophisticated arbitrary waveform generator, a more advanced algorithm could be used to produce arbitrary wave packets \cite{Shalibo2013}. As mentioned in the Introduction, the ability to produce shaped pulses controllably is very relevant for the good absorption of photon pulses by distant nodes of a quantum network.

\section{Photon generation efficiency}\label{sec:eff}
\begin{figure}[!hbt]
\centering
\includegraphics{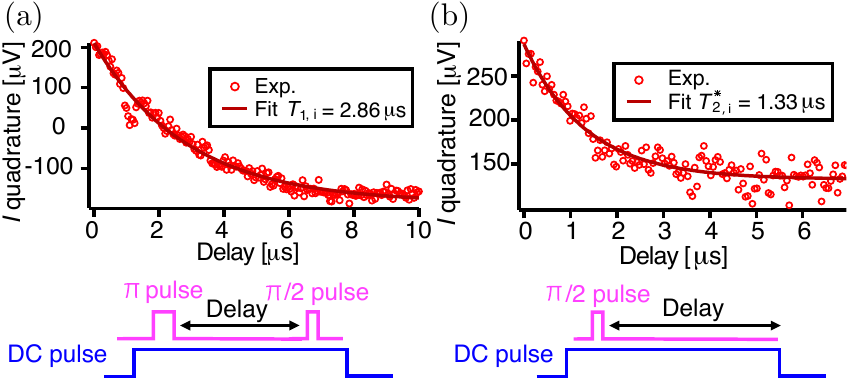}
\caption{\label{fig7}Intrinsic qubit coherence properties. (a) The intrinsic relaxation time $T_{1,i}$ and (b) the free induction decay time $T_{2,i}^{\ast}$ of the qubit are measured at the decoupling point following the pulse sequence indicated in the figure. (a) In order to measure $T_{1,i}$, a resonant $\pi$ pulse is delayed from a $\pi/2$ pulse that projects the qubit state onto the equatorial plane at the end of the sequence \cite{abdum2011}. When the SQUID current pulse ends the qubit spontaneously emits and its coherence properties are translated into the field quadrature amplitudes $I$ and $Q$. (b) The sequence to measure $T_{2,i}^{\ast}$ requires only a $\pi/2$ pulse that excites the qubit in a superposition state at the decoupling point followed by a time delay from the end of the SQUID current pulse which enables the qubit emission. The fact that $T_{2,i}^{\ast}<2T_{1,i}$ indicates the presence of a significant amount of intrinsic pure dephasing $\Gamma_{\phi,i}=(1/2.49)~\mu\rm{s}^{-1}$.}
\end{figure}

Finally, we can use our ability to control the photon emission to directly measure the intrinsic qubit coherence times which are necessary to calculate the photon generation efficiency. Adapting a technique developed at NEC \cite{abdum2011}, we first apply a $\pi$ pulse to the qubit at the decoupling point which excites it in the state $|1\rangle$. The qubit excitation decays with its intrinsic rate $\Gamma_{1,i} = 1/T_{1,i}$ due to the microscopic environment, which we refer to as nonradiative decay. This decay may include emission into other transmission lines, such as the excitation line, even though we estimate that channel loss to be negligible. Our observations cannot rule out that the qubit at the decoupling point has a remaining but small coupling to the source line, limiting the intrinsic relaxation time $T_{1,i}$. After a varying delay $\tau$, a $\pi/2$ pulse is applied to the qubit to rotate it onto the $\sigma_x-\sigma_y$ plane, followed by a current pulse which brings the qubit to a strongly emitting bias point. The $\pi/2$ pulse converts the remaining qubit population into a detectable field quadrature amplitude, as was shown in Fig.~\ref{fig3}. The integrated signal shown in Fig.~\ref{fig7}(a) for a varying delay leads us directly to an intrinsic relaxation time $T_{1,i}=2.86~\mu$s, which is a typical value for a transmon qubit with a nonoptimized geometry. 

Similarly, we can obtain the intrinsic dephasing time of the qubit $T_{2,i}^{\ast}$ by biasing it at the decoupling point and exciting it directly with a $\pi/2$ pulse, as shown in Fig.~\ref{fig7}(b). By controlling the delay time between the $\pi/2$ pulse and the end of the current pulse, we directly monitor the leftover qubit coherence. The integrated signal leads us to $T_{2,i}^{\ast} = 1.33~\mu$s, which is also in agreement with other experiments using planar transmon devices in resonators \cite{houck2007, bozyigit2011, pechal2014, kindel2016}. The obtained coherence times are compatible with a pure intrinsic qubit dephasing time of $T_{\phi,i} = 2.49~\mu\rm{s}$.

As explained in Sec.~\ref{sec6}, a figure of merit of an on-demand photon generator is the ratio of the storage time to the jitter.  For our device, we can take the intrinsic relaxation time, $T_{1,i}$, as the storage time and the minimum relaxation time, $T_{1,\,\rm{jit}}$, as the jitter. For this particular device, the shortest observed decoherence time is the value $T_{2}=81~\rm{ns}$ obtained in Sec.~\ref{sec6}, leading to $T_{1,\,\rm{jit}}=63~\rm{ns}$ using the dephasing value from Fig.~\ref{fig2}. Combined with the measured $T_{1,i}=2.86~\mu\rm{s}$ from Fig.~\ref{fig7}(a), we obtain a ratio of $T_{1,i}/T_{1,\,\rm jit}\approx47$. 

In order to estimate the efficiency of our circuit as a single-photon source, we perform a numerical calculation of the Lindblad master equation of a driven transmon using QuTip \cite{qutip} with the $T_{1,i}$ and $T_{\phi,i}$ values measured. We computed the efficiency of the source as the fidelity $\mathcal{F}$ to produce the targeted state $\rho$ given the generated state $\sigma$, $\mathcal{F} = \rm{Tr}\left[(\sqrt{\rho}\sigma\sqrt{\rho})^{1/2}\right]$. Our calculation also takes into account state preparation errors from leakage into higher transmon states \cite{Chow2010}, a finite temperature of $T_{\rm eff}=90~\rm{mK}$ determined from a separate experiment \cite{forn-diaz2017}, and parasitic decay into the excitation line. With a simple Gaussian preparation pulse, our estimated state preparation fidelity is 92\% for both $|\psi\rangle = |1\rangle$ and $|\psi\rangle = (|0\rangle + |1\rangle)/\sqrt{2}$. After a 100-ns storage time (the second lowest trace in Fig.~\ref{fig5}), the total efficiency of the generated propagating state, including decoherence during the emission time, is calculated to be 90\% for generating a Fock state $|1\rangle$ and 88\% for generating an equal superposition $(|0\rangle + |1\rangle)/\sqrt{2}$.  For a 1-$\mu\rm{s}$ storage time (the uppermost trace in Fig.~\ref{fig5}), the calculated efficiencies are 74\% and 73\% for $|\psi\rangle = |1\rangle$ and $|\psi\rangle = (|0\rangle + |1\rangle)/\sqrt{2}$, respectively.  We would like to emphasize that our source efficiency is limited neither by contamination from control photons nor decay during state preparation, as was the case in previous implementations \cite{houck2007, peng2016}.

As has already been shown possible \cite{barends2013}, the planar transmon circuit design can be engineered to enhance the values of the intrinsic coherence times in the tens of microseconds range. In addition, the qubit-line interaction strength can also be increased using alternative techniques from recent transmon-resonator experiments \cite{bosman2017}. 

\section{Conclusions}
In this work, we demonstrate the on-demand generation of single photons using a superconducting transmon qubit capacitively coupled to a transmission line. We show that the qubit-line interaction can be tuned by controlling the magnetic flux threading a dc SQUID, which acts as an inductive shunt at the end of the transmission line. The coupling can be tuned in real time, allowing the production of arbitrarily shaped single microwave photons with high fidelity. Our decoupling scheme can be directly implemented with other types of superconducting qubits, such as flux and fluxonium qubits, and also other solid-state qubits such as quantum dots \cite{qdots}. The technique developed in this work can be generalized to circuits with multiple qubits for the on-demand generation of many-body quantum states of radiation, including superradiant states. The decoupling scheme can be equally applied to resonators, potentially enabling the implementation of recent robust quantum communication protocols for microwave thermal fields \cite{vermersch-xiang2017}, complementing existing methods of controlled photon emission \cite{pfaff2017}.

\section*{Acknowledgements}
We acknowledge B.~L.~T.~Plourde for providing access to the evaporator at the University of Syracuse for fabrication of the aluminum Josephson circuits and J.~J.~Nelson and M.~Hutchings for their assistance. We also acknowledge our fruitful discussion with S.~R.~Sathyamoorthy and J.~Combes. We acknowledge NSERC of Canada, Industry Canada, the Canadian Foundation for Innovation, and the Ontario Ministry of Research and Innovation for funding. P.~F.-D. is currently supported by a Beatriu de Pin\'os fellowship.

%

\end{document}